\documentclass{aastex}          %% The manuscript based on AASTeX v5.x
\usepackage{spr-astr-addons}    %% mimicing ASTR journal style
\usepackage{url}
\usepackage{epstopdf}
\usepackage{hyperref}

\RequirePackage{color}

\newcommand{\emaila}{\url{daniel.gaslac-gallardo@unesp.br}}

\begin{document}

%% Article title
%
\title{Analysing the region of the rings and small satellites of Neptune}

%% Running heads
\shorttitle{Analysing the region of the rings and small satellites of Neptune}
\shortauthors{Gaslac Gallardo et al.}

%% Author and Affilations
\author{D.M. Gaslac Gallardo\altaffilmark{1}}
 \affil{Daniel M. Gaslac Gallardo}
 \affil{\emaila}
 \and \author{S.M.~Giuliatti Winter\altaffilmark{1}} \and \author{G.~Madeira\altaffilmark{1}} \and
 \author{M.A.~Mu\~noz-Guti\'errez\altaffilmark{2}}
%\email{\emaila} %% non-output

%% Alternate Affilations
\altaffiltext{1}{ Grupo de Din\^amica Orbital \& Planetologia, S\~ao Paulo State University-UNESP, Av. Ariberto Pereira da Cunha, 333, Guaratinguet\'a -SP, 12516-410, Brazil} 
\altaffiltext{2}{
Institute of Astronomy and Astrophysics, Academia Sinica, Taipei 10617, Taiwan}

%% Abstract
\begin{abstract}

The ring system and  small satellites of Neptune were discovered during Voyager~2 flyby in 1989 \citep{smith1989voyager}. In this work we analyse the diffusion maps which can  give an overview of the system.
As a result we found the width of  unstable and stable  regions close to each satellite.  
The innermost  Galle ring, { which} is further from the  { satellites, is} located in a stable region, while Lassel ring ($W= 4000$~km) has its inner border { in a} stable { region} depending on its eccentricity. The same happens to the Le Verrier and Adams rings{ , they are stable for small values of the eccentricity}. They can survive {  to the} close satellites perturbation only for { values of }   $e < 0.012$.
When the solar radiation force is { taken into account} the  rings composed  by $1\mu$m sized  particles { have} a lifetime { of} about $10^4$~years { while larger  particles  ($10\mu$m in radius) can survive up to $10^5$~years}.
The satellites Naiad, Thalassa and Despina can help replenish the lost particles of { the} Le Verrier, Arago and Lassel rings, while the ejecta produced by  Galatea, Larissa and Proteus { do} not have enough velocity to escape from the satellite gravity.

\end{abstract}

%% Keywords

\section{Introduction}
Six satellites of Neptune were discovered in 1989 during Voyager~2 flyby: Naiad, Thalassa, Despina, Galatea, Larissa and Proteus \citep{smith1989voyager}.  Proteus is larger compared to the other five satellites, which have radius smaller than 100km. Triton, the largest  satellite, and Nereid completed the Neptune satellite system until the discovery of Hippocamp, the smallest satellite (about 17km in radius, \cite{showalter2019seventh}).

Voyager 2 cameras also imaged a ring system formed by Galle, Le Verrier, Lassel, Arago and Adams rings
\citep{smith1989voyager}. Adams ring is a narrow ring composed by a sample of arcs named Courag\'e, Libert\'e, Fraternit\'e, Egalit\'e 1 and 2. These arcs have shown variations in brightness. Le Verrier and Arago rings are also narrow, while Lassel ring is larger and very faint. Galle ring is the innermost ring  { 2000km wide}. There is also a co-orbital ring with the satellite Galatea \citep{porco1995neptune}. 

Several papers have analysed the origin of the rings of Neptune \citep{colwell1992origins, colwell1993origins} and the history of
the inner small satellites \citep{banfield1992dynamical, zhang2008orbital}. It is also worth to analyse the system nowadays and the connection between the satellites, specially the satellites close   to the ring system.

The goal of this work is to analyse the region encompassing the orbits of  the small satellites and the ring system of Neptune, { as well as } the interaction among the satellites and the ring particles. The paper is divided into 4 sections. In section~2 we briefly describe the frequency map { analysis (FMA) technique applied}  for the satellites  region and the results regarding the diffusion maps. In section~3 we analyse  perturbative { forces, solar radiation force and plama drag, } acting in the ring particles and the r\^ole of each satellite in production ring material. In the last section we discuss our results.
%Your text comes here. Separate text sections with

\section{Satellites' region}

\label{sec:1}
%%%%%%%%%%%%%%%%%%%%%%%%%%%%%%%%%%%%%%%%%%%%%%%%%%%%%%%%%%%
%%%%%             Table 1
%%%%%%%%%%%%%%%%%%%%%%%%%%%%%%%%%%%%%%%%%%%%%%%%%%%%%%%%%%%
  \begin{table}[ht!]
  \centering
  \caption{Parameters of Neptune. Extracted from \cite{jacobson2009orbits}}
\label{tab:1}       % Give a unique label
\resizebox{7.0cm}{!}{
  \begin{tabular}{ccc}
   \hline
   Parameter &  \\  
   \hline
GM (km$^3$ s$^{-2}$)  & 6835099.502439672 $\pm$ 10 \\ 
$M_{N}$ ($\times10^{24}$)  &  102.41 $\pm$ 0.01 \\ 
R$_N$ (km)&  25225 &  \\
J$_2$ ($\times10^{-6}$)    & 3408.428530717952 $\pm$ 4.5 \\
J$_4$ ($\times10^{-6}$)    & -33.398917590066   $\pm$ 2.9\\
   \hline
   \end{tabular}
   }
   \end{table}
%%%%%%%%%%%%%%%%%%%%%%%%%%%%%%%%%%%%%%%%%%%%%%%%%%%%%%%%%%%
%%%%%             Table 2
%%%%%%%%%%%%%%%%%%%%%%%%%%%%%%%%%%%%%%%%%%%%%%%%%%%%%%%%%%%
%
% For tables use
\begin{table*}[ht!]
\centering
\caption{Parameters of the satellites}
\label{tab:2}       % Give a unique label
% For LaTeX tables use
\resizebox{14.63cm}{!}{
\begin{tabular}{lcccc}
\hline\noalign{\smallskip}
   Body & GM (km$^3$/s$^2$) & Mean Radius  (km) & Density (g/cm$^3$) & Reference\\  
\noalign{\smallskip}\hline\noalign{\smallskip}
Naiad   & 0.013  & 33 $\pm$ 3& 1.3 &  \cite{karkoschka2003sizes}\\
Thalassa& 0.025  & 41 $\pm$ 3& 1.3 &  \cite{karkoschka2003sizes}\\
Despina & 0.14   & 75 $\pm$ 3& 1.3 &  \cite{karkoschka2003sizes}\\ 
Galatea & 0.25   & 88 $\pm$ 4& 1.3 &  \cite{karkoschka2003sizes}\\
Larissa & 0.33   & 97 $\pm$ 3& 1.3 &  \cite{karkoschka2003sizes}\\
Hippocamp& 0.0003 & 17.4 $\pm$ 2 & 1.3 & \cite{showalter2013new,showalter2019seventh}\\ 
Proteus & 3.36   &210 $\pm$ 7& 1.3 &  \cite{karkoschka2003sizes}\\
Triton  &1427.598 $\pm$ 1.9& 1353.4 $\pm$ 0.9& 2.059 $\pm$ 0.005&  \cite{jacobson2009orbits,thomas2000shape}\\
\noalign{\smallskip}\hline
\end{tabular}
}
\end{table*}
%
%%%%%%%%%%%%%%%%%%%%%%%%%%%%%%%%%%%%%%%%%%%%%%%%%%%%%%%%%%%
%%%%%             Table 3
%%%%%%%%%%%%%%%%%%%%%%%%%%%%%%%%%%%%%%%%%%%%%%%%%%%%%%%%%%%
%
% For tables use
\begin{table}[ht!]
\centering
\caption{Parameters of Neptune's rings. Extracted from \cite{de2015planetary,de2019rings}}
\label{tab:3}       % Give a unique label
% For LaTeX tables use
\resizebox{8.33cm}{!}{
\begin{tabular}{lclcc}
\hline\noalign{\smallskip}
   Rings & Location (km) & $W$ (km)  & Optical Depth\\  
\noalign{\smallskip}\hline\noalign{\smallskip}

Galle      & 42000  & $\sim2000$ & $\sim10^{-4}$ \\
             
Le Verrier & 53200  & $\sim100$  & $3\times10^{-3}$  \\

Lassell    & 55200  & $\sim4000$ & $\sim10^{-4}$ \\ 

Arago      & 57200  & $\sim100$  &  ?  \\

Galatea co-orbital  & 61953    & $\sim50$ & ?  \\           

Adams      & 62933  & $\sim15$ & $\sim3\times10^{-3}$ \\

Adams { arcs} & 62933  & $\sim10$ &  $\sim10^{-1}$           \\
\noalign{\smallskip}\hline
\end{tabular}
}
\end{table}

%%%%%%%%%%%%%%%%%%%%%%%%%%%%%%%%%%%%%%%%%%%%%%%%%%%%%%%%%%%
%%%%%             Table 4
%%%%%%%%%%%%%%%%%%%%%%%%%%%%%%%%%%%%%%%%%%%%%%%%%%%%%%%%%%%
%
% For tables use
\begin{table}[ht!]
%\centering
\caption{Positions ($x$, $y$, $z$) and velocities ($v_x$, $v_y$, $v_z$) of the eight
satellites in  Neptune inner and Triton reference system (International 
Celestial Reference Frame - ICRF). The epoch for the state vectors is 
February~1,~2017 TDB. These values were extracted from { the} site web \footnotesize{HORIZONS}\tablenotemark{1}.}
\label{tab:4}       % Give a unique label
% For LaTeX tables use
\resizebox{8.415cm}{!}{
\begin{tabular}{lrr}
\hline
%\hline\noalign{\small}
   Satellites & Position (km)& Velocity (km/day)\\  
\hline
%\noalign{\smallskip}\hline\noalign{\small}
        &-4.861598101194755E+03&   -1.022607376451960E+06 \\% [2pt] % Naiad
Naiad   & 4.792192785052824E+04&   -1.007473952748110E+05\\%[2pt]
        &-2.513901855087704E+03&    5.737149709031385E+04\\ %[2pt]
  & & \\%[2pt]
        &-4.536127084395342E+04&   4.281143792543722E+05\\ %[2pt]% Thalassa
Thalassa&-2.123161142809179E+04&   -9.145989649730624E+05\\%[2pt] 
        &-6.882101766768756E+00&   -1.064699239230846E+04\\%[2pt]
  & & \\%[2pt]
        &3.691546376852450E+04 &  7.013763660435721E+05\\%[2pt] % Despina
Despina &-3.735597748047257E+04 &  6.933922389982601E+05 \\% [2pt]
        &-4.305177328186073E+02 &  5.299076528200922E+03\\%[2pt]
  & & \\%[2pt]
        &-6.092150479333453E+04 &   1.639911934009283E+05\\%[2pt]% Galatea
Galatea &-1.118656310246041E+04 &  -8.931195696477976E+05 \\%[2pt]
        &-1.351794468076787E+01 &  -8.423057548930225E+03\\%[2pt]
  & & \\%[2pt]
        & 3.220023154660776E+04  & -7.497834172900536E+05\\%[2pt]%   Larissa
Larissa & 6.607288323931834E+04  &  3.643338200205931E+05 \\% [2pt]
        & 7.839652855768873E+02  &  7.040661389542341E+03\\%[2pt]
  & & \\%[2pt]
         & 1.048292427174439E+05  & 6.050417340226601E+04\\%[2pt]%   Hipocamp
Hippocamp&-9.118392454374658E+03  & 6.939273288189896E+05 \\%[2pt] 
         &-4.486626166000306E+02  & 1.032242074417382E+04\\%[2pt]
  & & \\%[2pt]
         & 7.244449534921265E+04  & -5.191391052146224E+05\\%[2pt]  % Proteus
Proteus  &9.268331408207947E+04   & 4.052542585486157E+05 \\%[2pt] 
         &1.474307246998542E+03   & 8.462041124334502E+03\\%[2pt]
  & & \\%[2pt]
        &-3.384649833000187E+05  &  1.122533770419439E+05\\%[2pt]%   Triton
Triton  & 9.055438292073048E+04  &  3.353923148048871E+05\\%[2pt]
        & 5.562943106087373E+04  &  1.370100134831567E+05\\%[2pt]
%\noalign{\small}\hline
\hline
\end{tabular}
\tablenotetext{1}{https://ssd.jpl.nasa.gov/?horizons} }
\end{table}

In this section we analyse the region encompassing the orbits of  the small satellites and the rings  through the { FMA}
 technique. { The FMA is useful in identifying chaotic regions in dynamical systems } with arbitrary degrees of freedom. { FMA has been succesfully} 
applied for several dynamical systems \citep{laskar1990chaotic,laskar1993frequency,nesvorny1998three,robutel2001frequency,papaphilippou1998global}.

Following  \cite{munoz2017long} we construct { diffusion maps by performing a FMA of the quantity }  $z'(t) = a(t)~\exp(i~\lambda)$ for each { test} particle { in a grid of initial conditions covering the Neptune's satellites and rings region}. We applied the { frequency modified Fourier transform (FMFT)} algorithm \citep{vsidlichovsky1997frequency} over $z'(t)$ on { the adjacent time intervals} $[0,T]$ and $[T, 2T]$, where $T$ corresponds to half of the total { integration} time. { We} compare  the main frequencies $\nu_1$ and $\nu_2$ from each time interval {in order to calculate} the diffusion parameter $D$, defined  { following }  \cite{robutel2001frequency, correia2005coralie} { as}
$$D={{|\nu_1 - \nu_2|}\over{T}},$$

{ The $D$ parameter } gives { a} measure of the stability of the orbit of each particle{ , since} the value of 
$|\nu_1 - \nu_2|$ will be large for those particles in unstable orbits, { whereas} if this difference is small the particle will be in a stable orbit. The value of $D$ is proportional to the value of  $|\nu_1 - \nu_2|$. { To construct the diffusion maps}, the value of $\log D$ can be plotted in a {  diagram} $a \times e$, in  colour scale, for each initial condition of the particle.

We also calculated the diffusion time-scale ($t_D$) which is an estimation of the time required for a diffusion of the orbit of the particle in the radial direction. It can be given by
$t_D= (DP)^{-1}$, where $P$ is the period of the orbit of the particle.

The numerical integration was carried out through the Mercury package \citep{chambers1999} using the Burlish St\" oer 
integrator. The dynamical system is formed by Neptune, and its gravity coefficients $J_2$ and $J_4$, the  satellites 
Naiad, Thalassa, Despina, Galatea, Larissa, Hippocamp, Proteus and Triton and a sample of { thousands of} massless particles.
Table~\ref{tab:1} gives the physical parameters of Neptune:  mass ($M_N$), radius ($R_N$), and the gravity coefficients 
$J_2$ and $J_4$. Table~\ref{tab:2} gives the physical parameters of the eight satellites. All the simulations were 
perfomed at the Saturn Cluster of the Group of Planetology and Orbital Dynamics at UNESP. The { initial} positions and velocities 
of each satellite are given in Table~\ref{tab:4}. These values were taken from the Horizons website for the Julian Day 
2457785.5, which corresponds to the date February 1,~2017. 

In order to eliminate the short periodic variations due to 
the gravity coefficients of Neptune we transform the positions and velocities of the satellites  into the  geometric 
elements \citep{renner2006use}. The geometric elements  given in Table~\ref{tab:5} are the semimajor axis $a~ (\times 
D_{AR})$, where $D_{AR}$ is the semimajor axis of the Adams ring ($D_{AR}=62932.7$km (Table~\ref{tab:3})), the 
eccentricity $e$, the inclination $I$ in degrees, { and the angles} $\varpi$, $\Omega$ and $\lambda$ are the longitute of the pericentre, 
the longitude of the ascending node and the mean longitude, respectively. The parameters of the ring system,  semimajor 
axis in km, width ($W$) in km and optical depth, are described in Table~\ref{tab:3}. All other orbital elements { of the rings, i.e.} $e$, 
$I$, $\varpi$, $\Omega$ and $\lambda$ were taken as  zero. 

In order to produce the { diffusion maps} we performed numerical integrations of the system for a { total} time { of $10^4$ orbital periods of the most external satellite. In the range from $0.6004D_{AR}$ to $1.29D_{AR}$ the most external satellite is Larissa, therefore the numerical simulation lasted about 18~years. From  $1.3D_{AR}$ to $2.2009D_{AR}$ the most external satellite is Proteus, therefore the duration of the numerical integration was about 35~years.}

%%%%%%%%%%%%%%%%%%%%%%%%%%%%%%%%%%%%%%%%%%%%%%%%%%%%%%%%%%%
%%%%%             Table 5
%%%%%%%%%%%%%%%%%%%%%%%%%%%%%%%%%%%%%%%%%%%%%%%%%%%%%%%%%%%
\begin{table*}[ht!]
\centering
\caption{ Geometric elements of { the} inner satellites of Neptune system. These values were extracted from Table~\ref{tab:4} }
\label{tab:5}       % Give a unique label
% For LaTeX tables use
\resizebox{17.0cm}{!}{
\begin{tabular}{lcccrrrr}
%\hline\noalign{\smallskip}
   Satellite &$a(D_{AR})$ & $e$ & $I(^\circ)$ & $\varpi(^\circ)$ & $\Omega(^\circ)$ & $\lambda(^\circ)$\\ %  [2pt]
   \hline
& & & & & & & \\%[2pt] 
Naiad    & 0.7662936974& 0.0028933904& 4.3629736556& 188.2828786823& 138.9038344642&  96.0416928089\\%[2pt]
Thalassa & 0.7955775163& 0.0003268493& 0.6033164215&  20.7223081675&  24.3345566865& 205.0851889228 \\%[2pt]
Despina  & 0.8345261894& 0.0002964914& 0.5612815661&  50.2566688896&  11.4590451697& 314.6928820767 \\%[2pt]
Galatea  & 0.9842950420& 0.0000704953& 0.5311472156& 182.0518196399&   9.0559635089& 190.4038414171 \\%[2pt]
Larissa  & 1.1685233905& 0.0010692628& 0.7796604322& 129.5811402013&  12.4127449149&  64.1321648913 \\%[2pt]
Hippocamp& 1.6727265144& 0.0004180706& 0.8832101088& 344.0241374017&  11.0844491331& 355.0179103288 \\%[2pt]
Proteus  & 1.8691460666& 0.0003309937& 1.0283412166& 165.5275467689&   7.7072669384&  52.0269988864 \\%[2pt]
%\noalign{\smallskip}
\hline
\end{tabular}
}
\end{table*}
%%%%%%%%%%%%%%%%%%%%%%%%%%

{ A sample of test particles were distributed in a rectangular diagram, $a \times e$. The initial values  were adopted as follows: } i) the semimajor axis $a$ was taken in the range $0.6004D_{AR}$ to $2.2009D_{AR}$, with { a resolution} $\Delta a = 3 \times 10^{-4}$, ii) $0 \leq e \leq 0.04$, with $\Delta e = 4 \times 10^{-3}$, iii) the inclination $I$, longitude of the pericentre $\varpi$, longitude of the ascending node $\Omega$ and mean longitude $\lambda$  were assumed to be  zero. 

\begin{figure*}%[t]
\centering
\resizebox{1.9\columnwidth}{!}{%
  \includegraphics{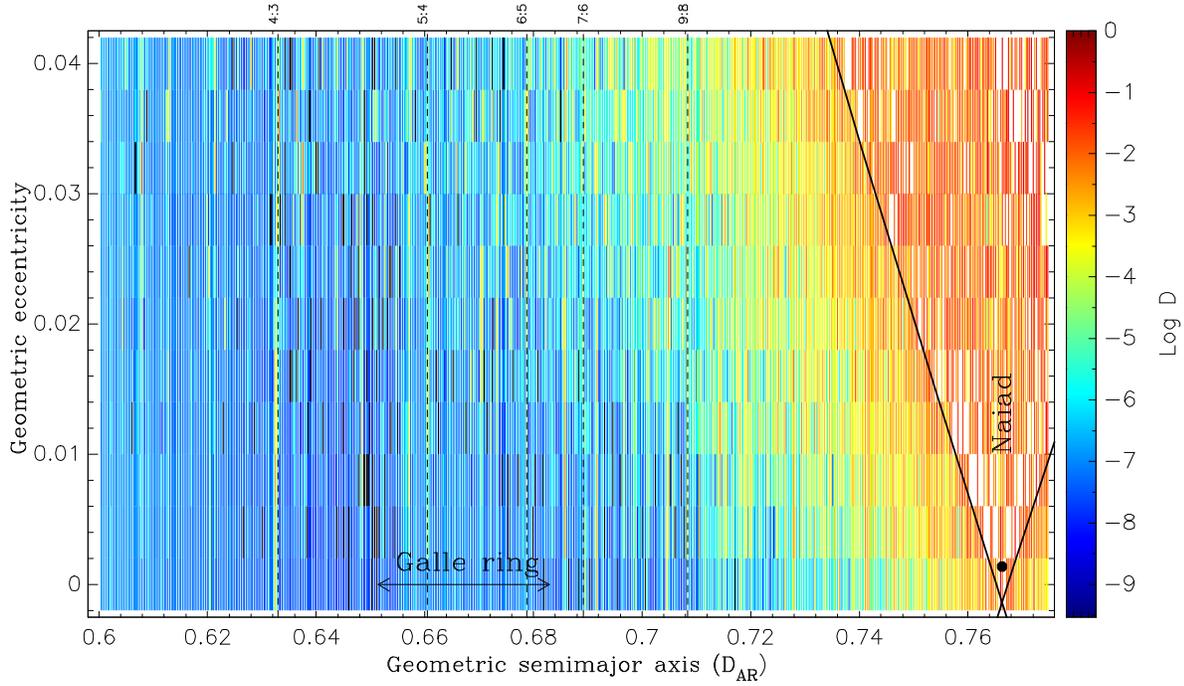} }
\caption{Diffusion map for  Naiad region. {  Galle ring} is located from 0.651$D_{AR}$ to $0.683D_{AR}$. The width of the ring can { is indicated } in the figure as a horizontal line. { Dashed black lines indicate } the location of the MMRs with Naiad. { Blue colours} correspond to stable regions, while { redder colours denote}  unstable regions }
\label{fig:1}       % Give a unique label
\end{figure*}

\begin{figure*}%[t]
\centering 
\resizebox{1.9\columnwidth}{!}{%
  \includegraphics{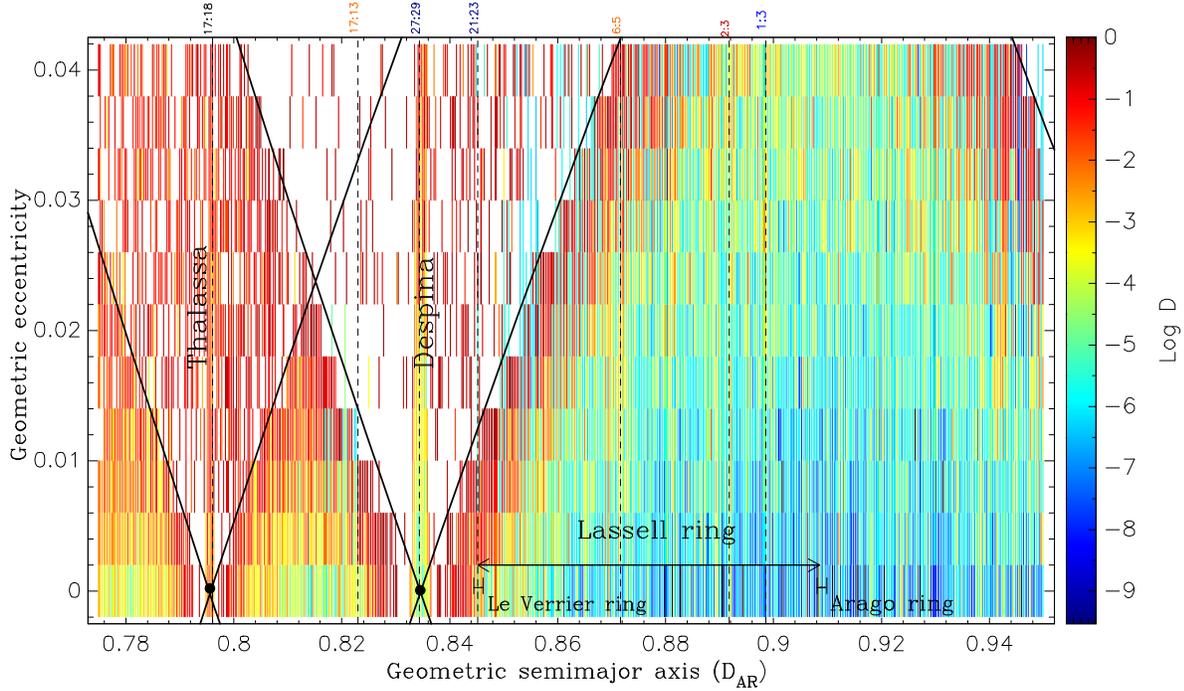} }
\caption{Diffusion map for  Thalassa and Despina { regions}. Despina cleared the unstable region.  
The MMRs with Naiad (in black colour), Thalassa (in dark blue), Galatea (in orange colour) and Larissa (in red colour) { are indicated at } the top of the figure}
\label{fig:2}       % Give a unique label
\end{figure*}

The { diffusion maps were} generated for  the region encompassing the orbits of the satellites  Naiad  to Proteus. The position of each satellite is identified by a black dot. In the following figures we describe each region in details.

Figure~1 shows the region where the Galle ring  and Naiad are located.  A { coloured} rectangle was plotted in the centre of each initial value of $a$ and $e$. { White} rectangles { indicate} that the particle collided or was ejected from the system. The { location and ratios of }
mean motion resonances (MMRs) with Naiad { are indicated with black dashed lines}  { and with labels at} the top of the figure.  The { solid} black { curves correspond} to the collision { curves}. The equations { defining such curves} are $a_s(1+ e_s) = a(1-e)$ if $a \geq a_s$ and  $a_s(1 - e_s) = a(1 + e)$ when $a \leq a_s$, where $a$, $e$, $a_s$, $e_s$ are the semimajor axis and eccentricity of the particle and the satellite, respectively. The values of the semimajor axis and  eccentricity of the satellites were obtained from Table~\ref{tab:5}.  Particles entering this region will cross the orbit of the satellite and will be { likely} ejected or will collide with the satellite. Most of the particles between these { curves} will collide with the satellite before the end of the numerical simulation. Since Naiad is a small { satellite, it  will} take longer time for clearing the region between these two { curves}. Galle ring is located  between  0.651$D_{AR}$ and $0.683D_{AR}$, in a stable region for almost any value of $e$.

\begin{figure*}%[t]
\centering
\resizebox{1.9\columnwidth}{!}{%
  \includegraphics{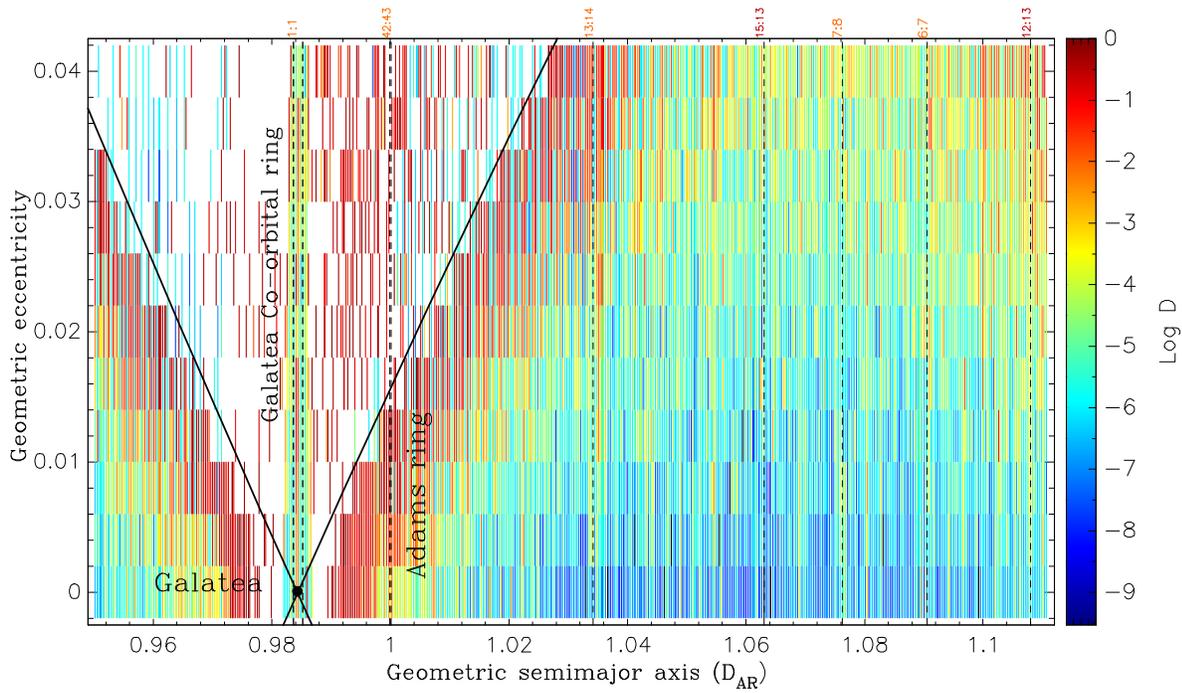} }
\caption{Diffusion map { for  Galatea} region. Two rings, the co-orbital ring of Galatea and the Adams ring, are { located} in this region }
\label{fig:3}       % Give a unique label
\end{figure*}

\begin{figure*}
\centering
\resizebox{1.9\columnwidth}{!}{%
  \includegraphics{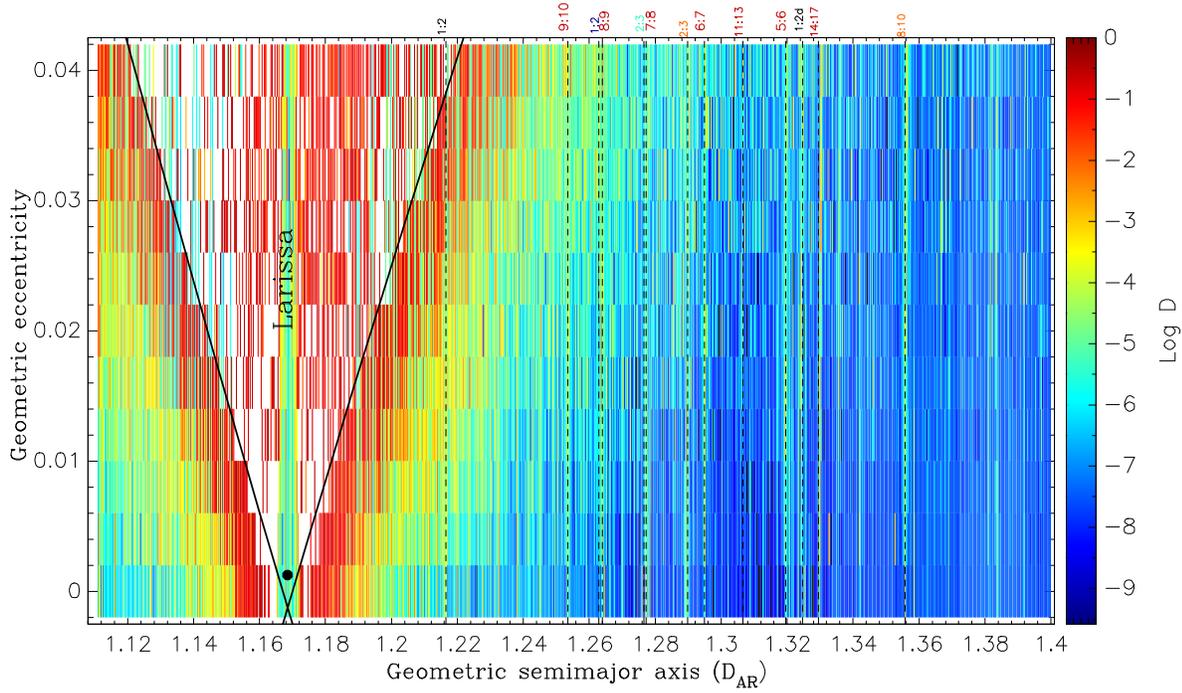} }
\caption{Diffusion map for  Larissa region. { A} stable region is { located} at 1.26$D_{AR}$-1.4$D_{AR}$}
\label{fig:4}       % Give a unique label
\end{figure*}

Thalassa and Despina cleared up the regions between the black { curves} in a short period of time (Figure~2). Most of the  region between these two satellites is unstable. Despina can hold a coorbital region for values of $e$ smaller than 0.01 (blue rectangles). The MMRs { ratios} with Naiad (in black colour), Thalassa (in dark blue), Galatea (in orange colour) and Larissa (in red colour) { are indicated } in the top of the figure. Thalassa and Larissa are close to the 17:18 MMR and Despina and Thalassa are close to the 27:29 MMR.  The inner edge of the  Lassel ring (0.845$D_{AR}$) is located in { an} unstable region overlapping the outer edge (0.846$D_{AR}$) of the Le Verrier ring. The same overlap occurs between the outer edge of the Lassel ring (0.908$D_{AR}$) and the inner edge of the Arago ring (0.908$D_{AR}$).
\begin{figure*}
\centering
\resizebox{1.9\columnwidth}{!}{%
 \includegraphics{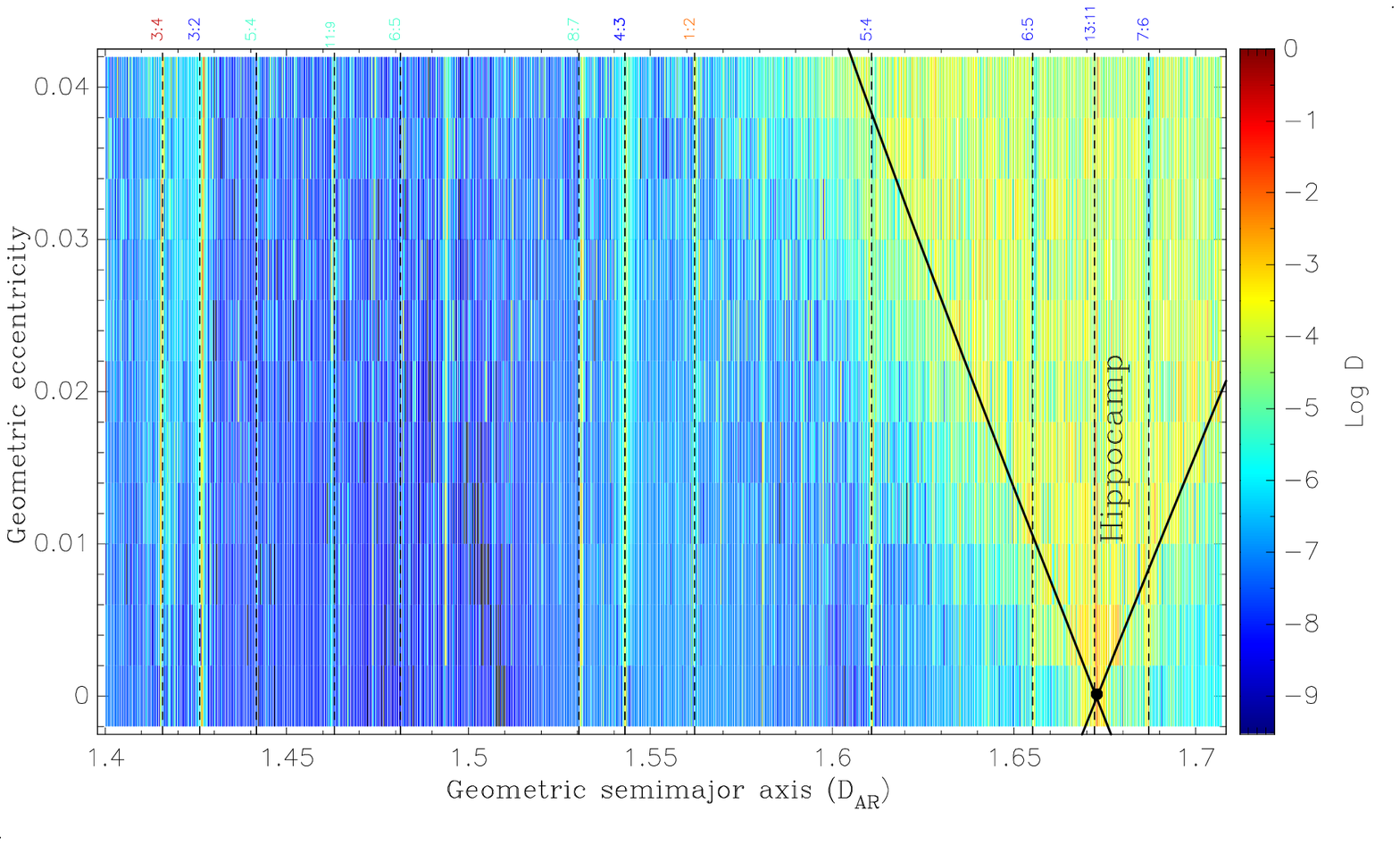} }
\caption{Diffusion map { for  Hippocamp region}. Due to the small effects of the satellite, this region is stable. Hippocamp can be in a 13:11 MMR with the satellite Proteus }
\label{fig:5}       % Give a unique label
\end{figure*}

\begin{figure*}
\centering
\resizebox{1.9\columnwidth}{!}{%
  \includegraphics{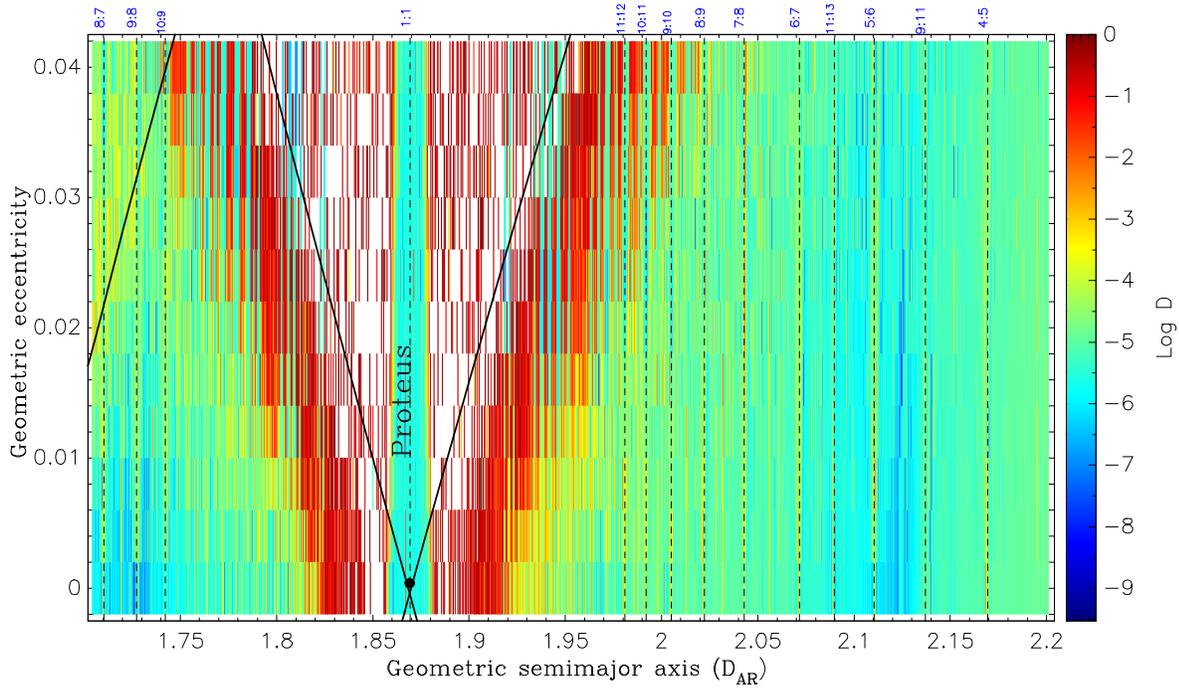} }
\caption{Diffusion map for Proteus region. All the MMRs shown in Figure~6 { are} related to the satellite Proteus }
\label{fig:6}       % Give a unique label
\end{figure*}

\begin{figure*}
\resizebox{1.9\columnwidth}{!}{%
  \includegraphics{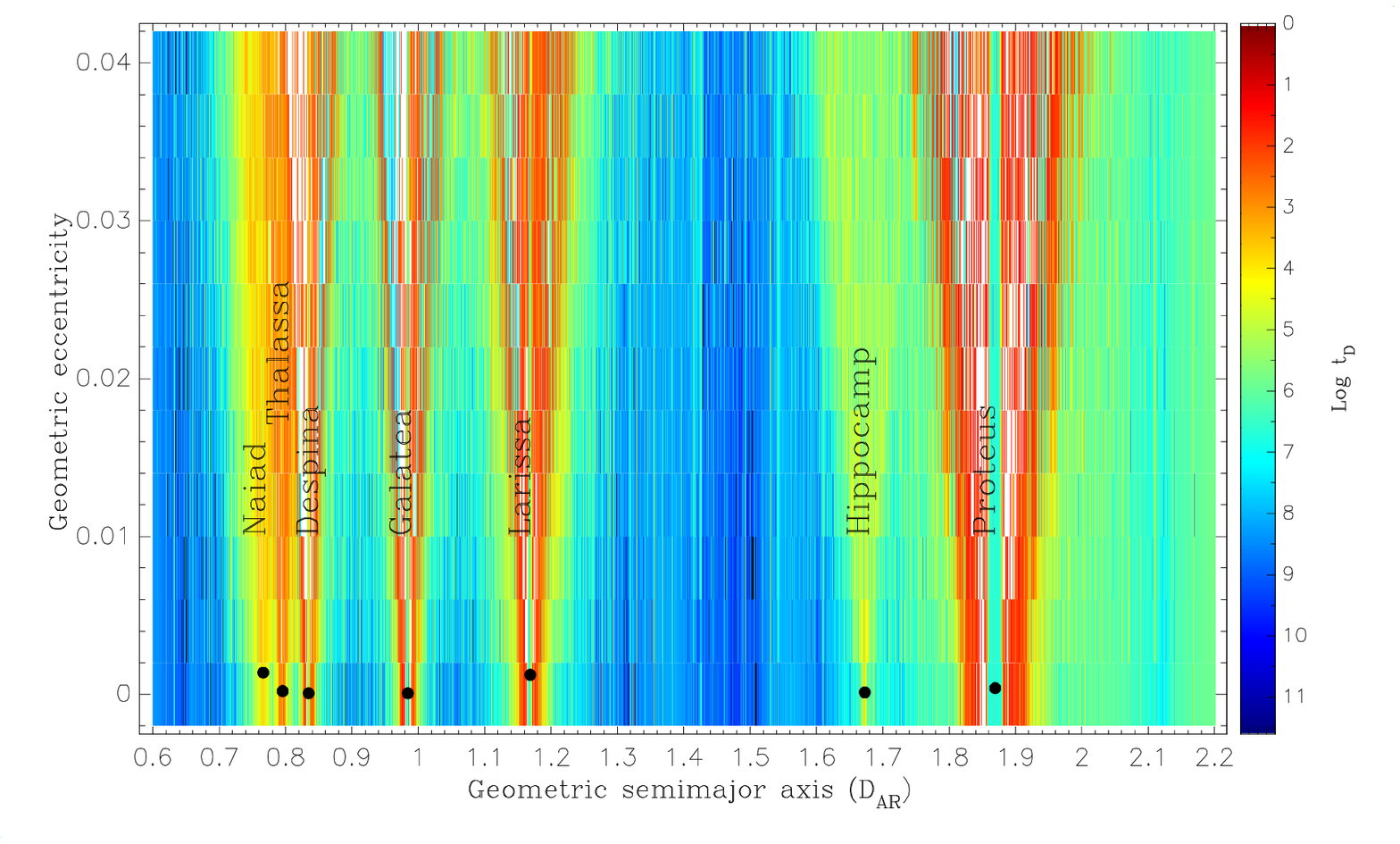} }
\caption{Diffusion time-scale for the  region. The satellites are { indicated } as black dots}
\label{fig:7}       % Give a unique label
\end{figure*}

Figure~\ref{fig:3} shows the diffusion map for the region surrounding the satellite Galatea, its co-orbital ring, Adams ring and its arcs. The Adams ring is located at 1$D_{AR}$. For small values of $e$ the ring particles are in a stable region. The { location of the } coorbital ring and Adams { ring are  shown} in { black} dashed lines. The MMRs are { shown at} { the} top of the figure{ ,  six} MMRs with Galatea (in orange colour) and two MMRs with the satellite Larissa (in red colour). As can be seen the Adams ring is close to the 42:43 MMR with Galatea, as proposed by the confinement mechanism model.

Larissa region is stable from 1.26$D_{AR}$--1.4$D_{AR}$ (Figure~\ref{fig:4}). The satellite clears the region between the diagonal black { curves} as expected. MMR's with Naiad (in black colour), Larissa (in red colour), Hippocamp (in green colour), Galatea (in orange colour) and Despina (1:2d) can be seen in the top of the figure.

Hippocamp is a very small satellite recently discovered by \cite{showalter2019seventh}. The region surrounding this satellite is stable. Several first and second orders MMR with Proteus, Hippocamp and Galatea can be seen in the top of Figure~\ref{fig:5}, one of this resonance is between Hippocamp and Proteus, 13:11~MMR. All the MMR's shown in Figure~\ref{fig:6} is with the satellite Proteus.  

\begin{figure*}[ht!]
\resizebox{1.0\columnwidth}{!}{%
  \includegraphics{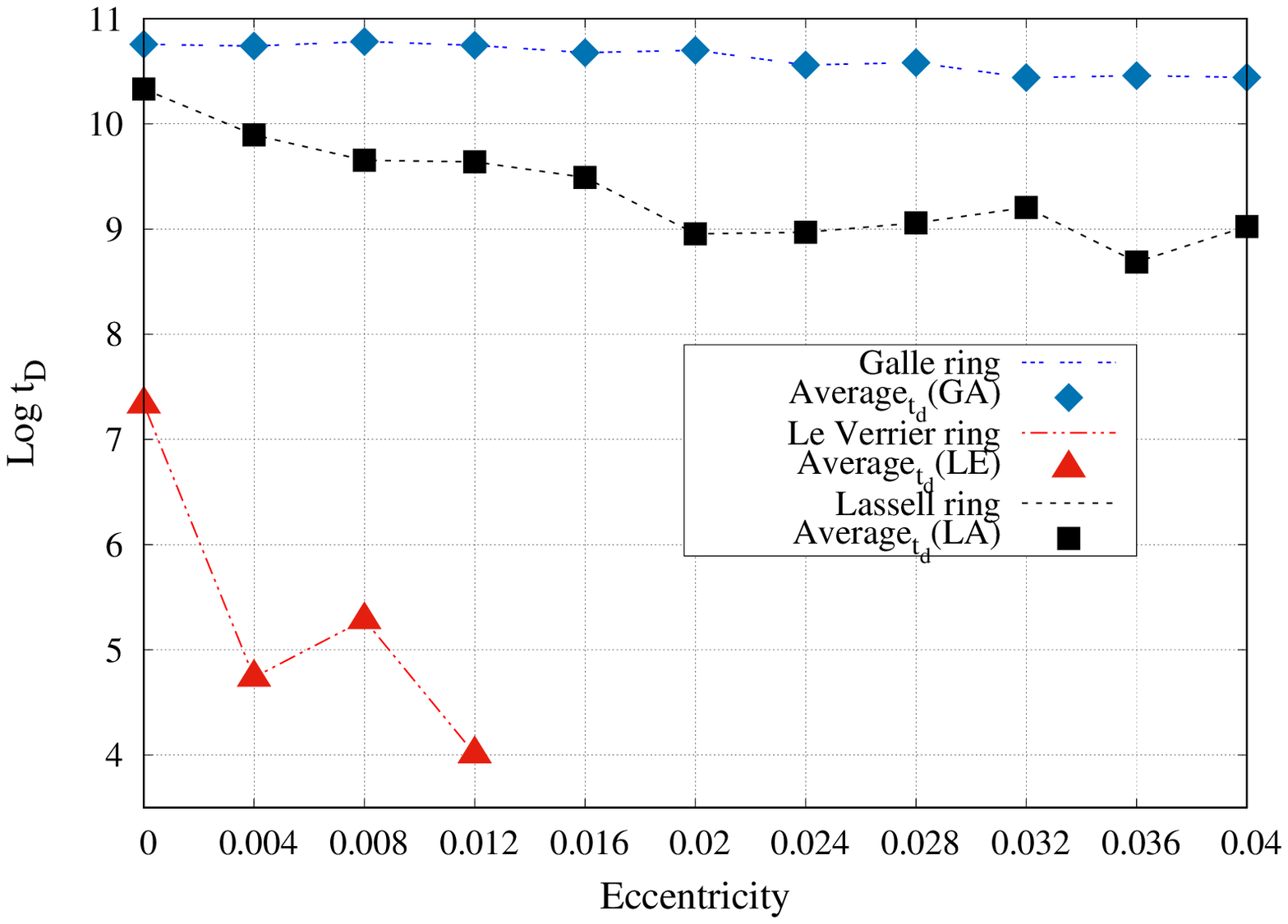}
  }
  \resizebox{1.0\columnwidth}{!}{%
  \includegraphics{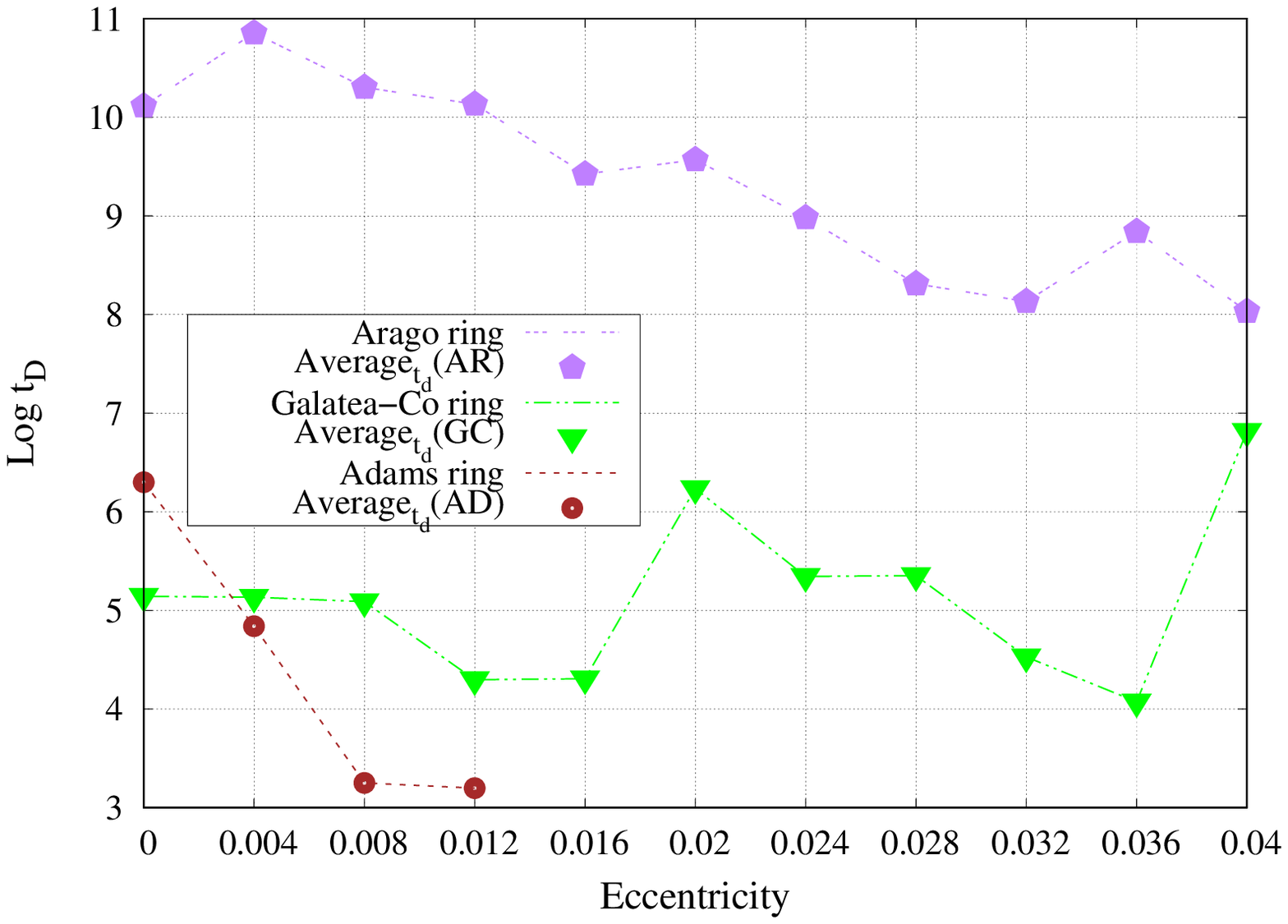}}
  %%\resizebox{0.75\columnwidth}{!}{%
  %%\includegraphics{fig8b.eps} }
\caption{Diffusion time-scale  of the rings: a) Galle, Le Verrier and Lassel rings and b) Arago, Galatea co-orbital   and  Adams rings }
\label{fig:8}       % Give a unique label
\end{figure*}

Figure~\ref{fig:7} shows the diffusion time-scale { map} for   { all the} region. { An} unstable region surrounds the satellite Proteus, the largest one. The diffusion time is about $10^2$~years. A large stable region can be seen between the orbits of the satellites Larissa and Hippocamp. The rings are also in stable regions (Figure~\ref{fig:8}), taking into account only the gravitational effects of the satellites. No dissipative forces { were included}  in the system up to now.  

Figure~8a shows the diffusion time-scale { ($t_D$)} for the  Galle, Le Verrier and Lassel { rings} for different values of the eccentricity.  The average { $t_D$} is relevant for the rings with large width. In fact it is relevant for the Lassel ring since it is located in different values of $log D$ (Figure~\ref{fig:2}). Galle ring is in a stable orbit with diffusion time of $10^{11}$~years, while the Le Verrier ring has an average diffusion time dependent on the eccentricity. This ring has a very short lifetime for values of $e = 0.012$. The Lassel ring has { $t_D$} about $10^9$~years.
Figure~8b shows the diffusion time-scale for the Arago, Galatea co-orbital and Adams rings. Arago ring has a diffusion time from $10^8$ to $10^{11}$ years depending on the eccentricity, while the Adams ring particles can survive up to $e \sim 0.012$, after this value of $e$ the ring particle enters the chaotic zone caused by the effects of  Galatea. Galatea co-orbital { ring} can survive up to about $10^5$~years.
%%
%\newpage

\section{Ring system under perturbative forces}

Regarding the {  dynamics of Neptune's rings} we showed  (section~2) that the rings are located  in stable positions, some { of them} are stable only  for small values of $e$.  However since these rings are composed of very tiny particles, these particles  suffer the effects of dissipative forces, such as the solar radiation force.  Figure~9 shows the strenght of the force  due to the oblateness of Neptune ($a_{\rm oblat}$), the solar radiation force ($a_{\rm SRF})$ and the gravitational effects of the satellite Triton ($a_{\rm T}$) on a $1\mu$m sized particle. These equations are given by \citep{murray1999solar, mignard1984effects}:
$$ a_{\rm oblat}= {{3GM_N J_2 R_N^2}\over{a^4}}  $$ 
$$ a_{\rm SRF} = {{3 \Phi Q_{\rm pr}}\over {4c \rho r}}$$
$$a_{\rm T} = {{G m_T}\over{(a-a_T)}}$$
where  $\Phi$ is the solar flux, $Q_{\rm pr} $ is a constant taken as 1 \citep{mignard1984effects}, $c$ is the speed of  light, $\rho$ is the density of the particle (assumed to be 1g/cm$^3$), $r$ is the radius of the particle,
 $m_T$ and $a_T$ are the mass and semimajor axis of Triton, respectively.
 
 As can be seen in Figure~9 the solar radiation force produces less effects on the particles  compared to the effects of $J_2$  and the gravitational effects of Triton.

 \begin{figure}[t]
\resizebox{0.95\columnwidth}{!}{%
  \includegraphics{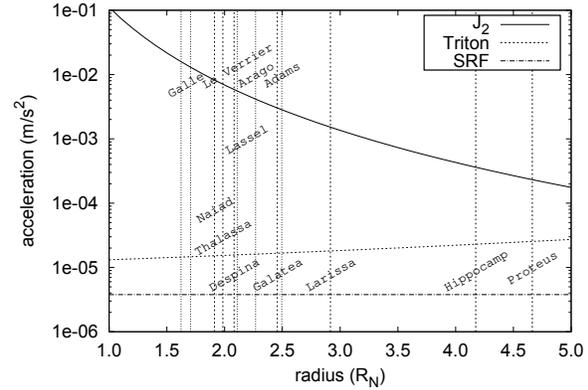} 
 }
\caption{Strenght of the forces in Neptune ring system as { a } function of { the} distance. The locations of the satellites and the rings are 
displayed  as vertical lines }
\label{fig:9}       % Give a unique label
\end{figure}

 In order to analyse the effects of this { dissipative} force a sample of numerical simulations of the ring particles under the  effects of $J_2$ and $J_4$ of Neptune and all the eight satellites  was carried out.
   Figure~10 shows the time variation of the eccentricity of a representative particle located in each ring, Galle, Le Verrier, Lassel, Arago and Adams { rings}.  
 
The eccentricity of a representative 1$\mu$m sized particle increases due to the effects of the solar radiation force. However  this increase is small (Figure~10). All the eccentricities remain of order $10^{-3}$.  There are two plots related to the Adams ring, the plot labelled Adams-LER 42:43 means that the ring particle is located in the resonance with the satellite Galatea. The confinement due to this resonance reduces the increase in the eccentricity of the ring particle. Adams ring and its arcs are under investigation by  Giuliatti~Winter, Madeira \& Sfair~2019. 
 
 Due to the { increase} in the eccentricity, the ring particles  can alter their radial excursions, but despite of that they will not cross the orbits of any satellite. The edges of  Le Verrier, Lassel and Arago { rings} will have the overlap increased.

\begin{figure*}[t]
\centering
\resizebox{2.07\columnwidth}{!}{%
\includegraphics{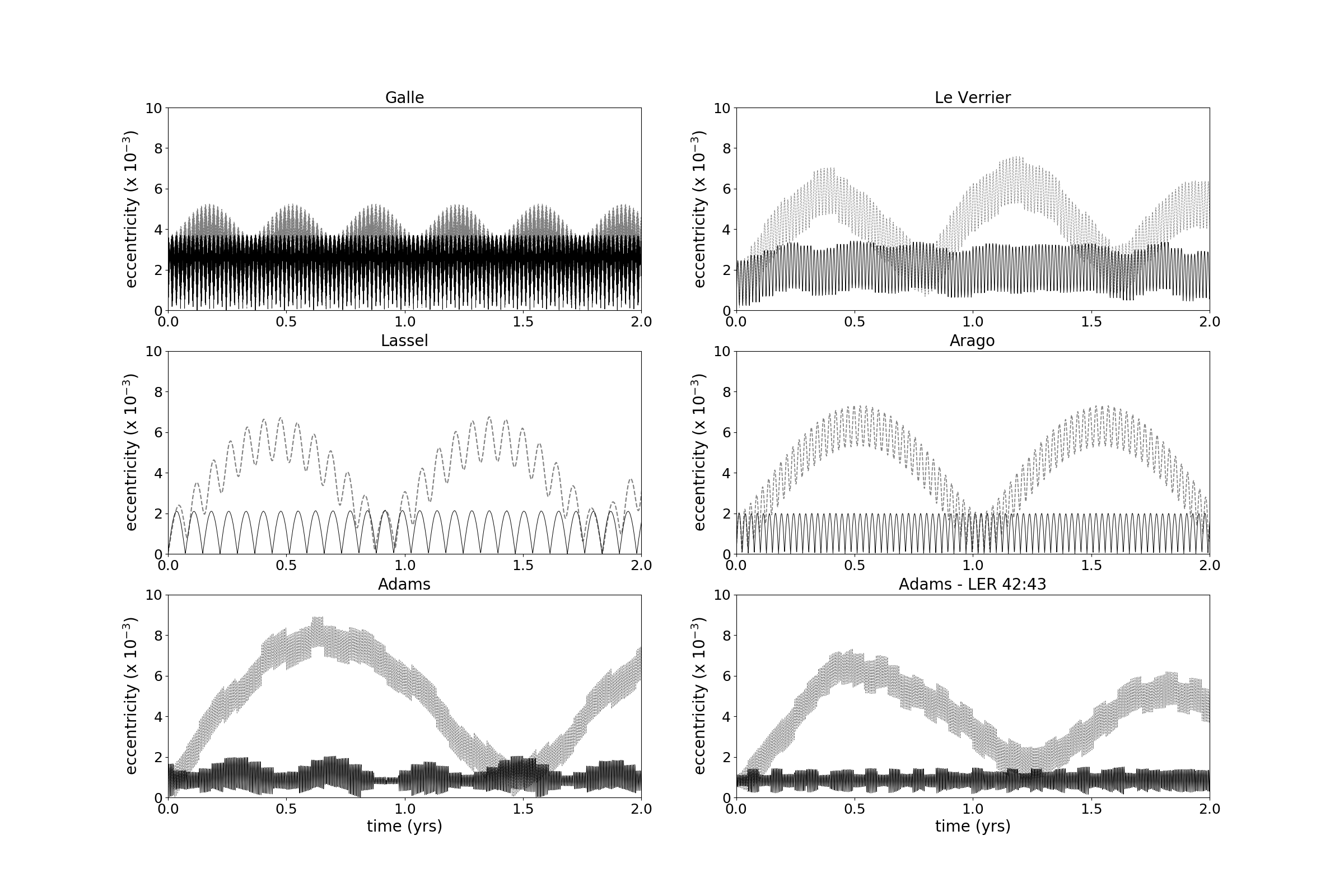} }
\caption{Time variation of the eccentricity of a representative  particle of each ring. The name of the ring is shown in the top of each figure. Each plot has two curves, without and with the effects of the solar radiation pressure. The effects of { this dissipative force}  is responsible for the small increase in the eccentricitiy of the 1$\mu$m~sized particle}
\label{fig:10}       % Give a unique label
\end{figure*}

Concerning the co-orbital ring to Galatea our results showed that although all
 the small particles remained azimuthally confined (no collision with Galatea was detected), the
smaller particle (1$\mu$m in radius) performs radial oscillations of about 900~km, while the oscillation
of a particle of 10$\mu$m in radius is only $\sim$ 120~km. These radial excursions do not cross the inner
edge of the Adams ring. Larger particles (30$\mu$m, 50$\mu$m and 100$\mu$m in radius), as expected, stay close to the
co-orbital region. In $10^5$~years the semimajor axis of the smaller particle decreases up to 7000~km,
crossing the Arago ring and the outer edge of the Lassel ring (57200~km). 

The timescale for  decay of the tiny particles  caused by the Poynting-Robertson component  gives  the time each ring will cross the orbit of the satellite and therefore its lifetime. The Poynting-Robertson component of the solar radiation force provokes a secular  decrease in the semimajor axis of the particle and the timescale ($\tau_{\rm pr}$) can by given by \citep{mignard1984effects}
$$\tau_{\rm pr} \sim 10^6~r  $$

{ Galle  1$\mu$m sized ring particles will collide with Neptune in $5 \times 10^5$~years while larger particles (10$\mu$m in radius) will take $5 \times 10^6$~years to reach Neptune.} Le Verrier, Lassel, Arago and Adams { rings} have similar lifetimes, small particles (1$\mu$m in radius) will reach the orbit of a given satellite in $10^4$~years   and larger particles (10$\mu$m in radius) in $10^5$~years.
These values  are given in Table~6.

\begin{table*}[ht!]
\centering
\caption{Lifetime of the rings particles (1$\mu$m and 10$\mu$m in radius) due to Poynting-Robertson effects}
\label{lifetime}       % Give a unique label
% For LaTeX tables use
\begin{tabular}{lcccc}
\hline\noalign{\smallskip}
ring & 	t ($10^4$ years) - $1~\mu m$ & t ($10^6$ years) - $10~\mu m$ & cross the orbit of  \\
\noalign{\smallskip}\hline\noalign{\smallskip}
Galle & 50 & 5 & { collide with Neptune}\\
Le Verrier & 1.1 & 0.11 & Despina\\
Lassel & 4.8 & 0.48 & Despina\\
Arago & 8.5 & 0.85 & Despina\\
Adams & 1.6 & 0.16 & Galatea\\
\noalign{\smallskip}\hline
\end{tabular}
\end{table*}

{ 

Plasma drag can also reduce the lifetime of the ring particles, however, we did not include these force due to few information on the Neptune plasmasphere. The azimuthal force caused by the plasmasphere's ions in the ring particles is given by \citep{hedman2013horseshoes}
$$ F_{PD}=\pi a^2(n-n_N)^2r^2\rho_i $$
where $n_N$ is Neptune's rotation rate\break ($n_N\sim 10.9$~rad/day) and $\rho_i$ is the plasma ion mass density.
The plasma drag effect increases the semimajor axis of the particles in the rings region. The estimated time variation of the semimajor axis is
$$ \frac{da}{dt} \sim 10^{-2}\left(\frac{\rho_i}{1~{\rm amu}/{\rm cm}^3}\right)\left(\frac{1~\mu {\rm m}}{r}\right)~{\rm km/year} $$
A very rough estimative was performed by assuming the plasmasphere  as composed primarily of $H^+$ (1~amu) and number density as $10^3/{\rm cm}^3$ \citep{lyons1995metal}. We obtained that the semimajor axis of $1~\mu$m sized particles increases of order of 10~km per year. The time to the ring particles cross the orbit of the outer satellite is given in Table~\ref{lifepd}.}

\begin{table*}[ht!]
\centering
\caption{Lifetime of the rings particles (1$\mu$m and 10$\mu$m in radius) due to plasma drag effects}
\label{lifepd}       % Give a unique label
% For LaTeX tables use
\begin{tabular}{lcccc}
\hline\noalign{\smallskip}
ring & 	t (years) - $1~\mu m$ & t (years) - $10~\mu m$ & cross the orbit of  \\
\noalign{\smallskip}\hline\noalign{\smallskip}
Galle & 400 & 4000 & Naiad\\
Le Verrier & 800 & 8000 & Galatea\\
Lassel & 700 & 7000 & Galatea\\
Arago & 600 & 6000 & Galatea\\
Adams & 1400 & 14000 & Larissa\\
\noalign{\smallskip}\hline
\end{tabular}
\end{table*}

\vskip20pt

We can also analyse the r\^ole of each satellite in producing ring material. A satellite can { produce} material due to impacts of interplanetary dust particles (IDPs) onto its surface. The material ejected can feed a tenuos ring.  By assuming  the impactors mass flux at Neptune's region as $F_{imp}^{\infty}=10^{-17}$~kg/(m$^2$s) and the mean velocity of the impactors as $v_{imp}^{\infty}=3.0$~km/s \citep{poppe2016improved}, we determined the mass flux $F_{imp}$ and velocity $v_{imp}$ of impactors enhanced by to the gravitational focusing of the planet through the algorithm described in Sfair and Giuliatti~Winter~2012 and  \cite{madeira2018production}.
The mass production rate by an icy moon with radius $R$ is \citep{koschny2001impacts, krivov2003impact} 
$$M^+=1.2\times 10^{-6} R^2F_{imp} v_{imp}^{2.46}$$

\begin{figure}[t]
\centering
\resizebox{0.9\columnwidth}{!}{%
\includegraphics{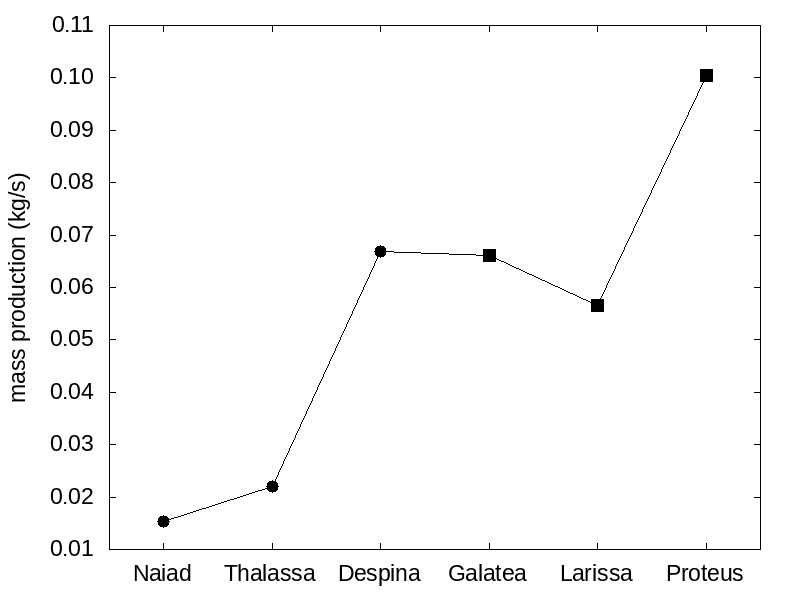} }
\caption{Mass production rate of the neptunian inner satellites}
\label{fig:10}       % Give a unique label
\end{figure}

Figure~11 shows the mass production rate of the neptunian inner satellites Naiad, Thalassa, Despina, Galatea, Larissa and Proteus. The squares { refer} to the satellites whose escape velocity is higher than the velocity of the dust particles, while the circles are for  the satellites whose escape velocity is smaller than the ejecta velocity. Therefore Galatea, Larissa and Proteus can produce dust particles{ , although} most of the grains can not  leave the satellites surface. {  The result regarding the satellite Galatea is derived from \cite{giuliattiwinter2019orbital}.}

A crude estimative can be made by comparing the mass of the ring and the mass  production of a satellite immersed on it. The mass of  Neptune ring can be calculated using the algorithm presented in Sfair and Giuliatti~Winter~2012. Assuming $\tau = 3.3 \times 10^{-3}$ for the  Le Verrier and  Arago { rings} their masses are about 
$5 \times 10^8$~kg and $4.5 \times 10^8$~kg, respectively. Lassel ring mass is about $6 \times 10^8$~kg. 
Naiad, Thalassa and Despina can produce dust particles and populate these rings in less than 1000~years. However, these satellites are not immersed in these rings and part of these particles will be lost until { they can} reach the ring. Nevertheless Naiad, Thalassa and  Despina may contribute to the ring population.

%\vskip0pt

\section{Discussion}

The diffusion maps give an overview of the system populated by small satellites and ring particles of Neptune. The largest unstable region is surrouding  Proteus, the large satellite. Between the satellites Larissa and Hippocamp is located the largest stable region. Several MMRs of first and second orders were identified, as examples, Thalassa and Larissa are close to the 17:18 MMR, Despina and Thalassa are close to 27:29 MMR and the recent discovery satellite Hippocamp is in the 13:11 MMR with Proteus.  

When only gravitational force is present { in the system} most of the rings are located in stable regions. The innermost  Galle ring is further from the satellites and is located in a stable region, while Lassel ring ($W= 4000$~km) has its inner border stable depending on its eccentricity. The same occurs to the Le Verrier ring, this ring can survive the perturbations caused by Despina only if its eccentricitiy is smaller than 0.012.  Adams ring is also stable for values of  $e < 0.012$.

Solar radiation pressure is important for the ring particles of Jupiter, Saturn, Uranus and even Pluto. In Neptune system this dissipative force  causes only very small variations in the eccentricities of the ring particles. 
The eccentrities of the ring particles are of order  $10^{-3}$ due to the gravitational perturbations of the 
satellites and  stay at this same order even when the solar radiation force is taken into account. The particles do not cross the orbits of the satellites  when these new values of eccentricity are assumed.

The lifetime of the rings particles is probably dictate by the decreasing in the semimajor axis due to secular effects of the Poynting-Robertson component of the solar radiation force. Although the lifetime caused by the perturbations of the nearby satellites is close to this value (Figure~8). The lifetime of the rings composed  by $1\mu$m sized  particles is about $10^4$~years and about $10^5$ when the particles are larger ($10\mu$m in radius). {  The plasma drag may be an important dissipative force but we need further information to constrain its effects in the ring particles.}

The satellites Naiad, Thalassa and Despina can help replenish the lost particles of  Le Verrier, Arago and Lassel rings. The velocities of the ejecta particles are larger than the escape velocities and the ejecta particles can reach the rings. However, the satellites Galatea { \citep{giuliattiwinter2019orbital}}, Larissa and Proteus also produce ejecta material, but { they do} not have enough velocity to escape from the satellite gravity.

{ A paper by \cite{brozovic2019orbits} was published during the revision of this work. \cite{brozovic2019orbits} derived  orbital fits for the small inner satellites of Neptune based on data obtained by telescopes, Voyager~2 and the Hubble Space Telescope. They found that the satellites Naiad and Thalassa are in a 73:69 inclination resonance, while Hippocamp and Proteus are in a  13:11 near mean motion resonance. }

{ A detailed analysis of these probably resonances between the satellites and the satellites and  ring particles are under investigation.}

%% Acknowledgements
%
\acknowledgments
% <Acnowledgments text>
 The authors  thank Fapesp\break (2016/24488-0, 2016/24561-0 and 2018/23568-6) and CNPq (309714/ 2016-
8) for the financial support. This study was financed in part by the Coordena\c c\~ao de
Aperfei\c coamento de Pessoal de N\'\i vel Superior - Brasil (CAPES) - Finance Code~001.
%
%% References
%% Please cite all reference entries in the article text using \cite or
%% equivalent command. 
%%%  Using BibTeX  (Name-Year style)
%
 \bibliographystyle{spr-mp-nameyear-cnd}  %% BibTeX style
 \bibliography{manuscript}                %% BibTeX data
%% Non-BibTeX  (Name-Year style)
%
% \begin{thebibliography}{}
% \bibitem[\protect\citeauthoryear{<author>}{<year>]{ref:?}
%    <ref. entry>
% \bibitem[\protect\citeauthoryear{<author>}{<year>]{ref:?}
%    <ref. entry>
% \end{thebibliography}

\end{document}